# Confinement by Carbon Nanotubes Drastically Alters the Boiling and Critical Behavior of Water Droplets

Vitaly V. Chaban[1], Victor V. Prezhdo[2] and Oleg V. Prezhdo[1,*]

[1]*Department of Chemistry, University of Rochester, Rochester, NY 14627, USA*

[2]*Institute of Chemistry, Jan Kochanowski University, 25-406 Kielce, Poland*

Vapor pressure grows rapidly above the boiling temperature, and past the critical point liquid droplets disintegrate. Our atomistic simulations show that this sequence of events is reversed inside carbon nanotubes (CNT). Droplets disintegrate first and at low temperature, while pressure remains small. The droplet disintegration temperature is independent of the CNT diameter. In contrast, depending on CNT diameter, a temperature that is much higher than the bulk boiling temperature is required to raise the internal pressure. The control over pressure by CNT size can be useful for therapeutic drug delivery.

**Keywords:** phase transition, carbon nanotube, water, atomistic simulation

[*] Corresponding author, email: oleg.prezhdo@rochester.edu; v.chaban@rochester.edu; vvchaban@gmail.com

TOC Graphic

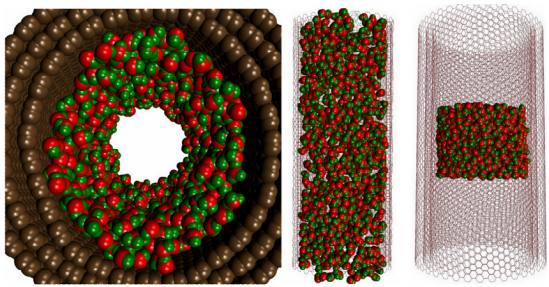

Confinement of conventional liquids inside nanoscale structures gives rise to fascinating physical properties that form the basis for an ever increasing number of novel applications. Significant efforts have been devoted to the study of structure, dynamics and thermodynamics of water inside carbon nanotubes (CNTs).[1-8] Both experimental and computational studies suggest that confined fluids exhibit a variety of new properties that differ greatly from those of bulk fluids. For instance, water readily forms ice-like structures inside CNTs even at room temperature.[4,9] Confinement notably softens water dynamics, resulting in a drastic reduction of shear viscosity as the CNT diameter decreases to a few effective solvent diameters.[10] Polar solute molecules, driven by dipole-dipole interactions, tend to aggregate,[5] and the aggregation is essential for CNT filling.

Thermodynamics of solid/liquid phase transitions has received significant attention due to important implications in biology and materials science. Hydrogen-bonded chains of water structures in protein ion channels can greatly facilitate proton transport across cellular membranes. Uniform diffusion of ionic solutions through nanoscale carbon architectures is essential for reliable charging and discharging of supercapacitors.[11] Investigations show that freezing under confinement can take place by both first-order and continuous phase transitions,[3,12] while bulk liquids freeze only through first-order transitions. Liquid/vapor coexistence of confined fluids has received significantly less attention. It has been found for n-alkanes that the critical point exists at lower temperature and higher density, compared to bulk.[13] This result can be explained by competition between fluid-fluid and fluid-surface interactions in combination with the entropic effect.

CNTs are promising candidates for novel therapeutic technologies.[14-16] Their unique mechanical, electrical, and optical properties can be utilized simultaneously to create drug delivery vehicles.[5] Hydrophobic CNTs penetrate through cell membranes and reach locations that are generally inaccessible to polar drug molecules.[17,18] Hence, one can encapsulate drugs inside CNTs and deliver them to desired parts of the living body. Unlike living tissues, CNTs absorb near-infrared light, which allows for the photoactivation of *in vivo* CNTs without damaging side effects.[19,20] Upon activation, the photon energy rapidly transfers from electrons to phonons[21] and, subsequently, is released as heat to the encapsulated solution.[22] Heating destroys CNT capping agents, freeing the drug. Increasing the vapor pressure of the encapsulated solvent greatly contributes to the drug release process.[23]

The current paper establishes the effects of spatial confinement on the evaporation and resulting vapor pressure of water droplets encapsulated within CNTs. Using classical molecular dynamics (MD), we show that vaporization of water inside CNTs occurs by an entirely different mechanism than in the bulk. In the bulk phase, boiling is followed by droplet disintegration, as the temperature rises above the critical point. In contrast, water droplets inside CNTs disintegrate at a significantly lower temperature and pressure than free droplets and very low pressure. In order to achieve the rapid pressure growth associated with boiling, confined water must be heated to a temperature far beyond the droplet disintegration temperature. While the temperature required for the pressure growth depends on the CNT diameter, the droplet disintegration temperature is diameter independent. The latter fact is quite remarkable, since most properties of confined liquids do depend on the CNT size. We show that the mechanism of water droplet evaporation inside CNTs is controlled by the strength of the CNT-water interaction.

## RESULTS AND DISCUSSION

**Bulk Liquid/Vapor Interface.** Liquid vaporization can be simulated by introducing a liquid/vapor interface in the simulation box, Figure 1. This can be achieved starting from conventional bulk liquid and extending the box along one of its three Cartesian directions. The resulting system contains an endless film of liquid surrounded by several nanometers of vacuum. To simulate the interface, it is critical that both liquid and gas phases are present simultaneously in the simulation cell. Since generally, vapor is three orders of magnitude less dense than liquid, proper representation of the gas phase requires a significantly larger simulation volume compared to the liquid phase.

In order to select an optimal box size for the simulation of the liquid/vapor interface, we tested the thermodynamic properties of three systems with different simulation volume and the same amount of liquid, Table 1. The liquid film was centered in boxes with the dimensions of 3.74 x 3.74 x 10, 3.74 x 3.74 x 20 and 3.74 x 3.74 x 30 nm. The internal pressure in these systems is generated mostly by the gas phase, and hence, the pressure component normal to the film surface is dominant. Water boils when its vapor pressure exceeds one atmosphere (for standard conditions). For TIP4P water, this happens at 368 K,[23] which is close to the experimental value of 373 K. The simulated boiling temperature is independent of the simulation box volume, as should be expected, Figure 2.

The three curves shown in Figure 2 diverge at the critical point of TIP4P water. At this temperature the liquid layer of water disappears, and pressure becomes dependent on the simulation box volume. The TIP4P model of water underestimates the experimental values of the critical temperature and pressure, 647 K and 228.5 bar, respectively. The improved TIP4P/2005 model agrees better with experiment in this case.[24] Since both boiling and critical points coincide for the three MD simulation cells, we concluded that all of them provide proper representation of the liquid/vapor interface. In order to save computer resources, we chose the smallest box for the production simulations.

**Disintegration of Confined Water Droplets and Anomalous Pressure Changes.** Next, we investigated pressure changes inside the (11,11), (21,21), (31,31) CNTs, Table 1, induced by gradual heating in the constant temperature constant volume ensemble. The resulting curves shown in Figure 3 are qualitatively different from the data for bulk water. First, the temperature required for a rapid pressure growth depends on the CNT diameter, and second, the initial pressure growth is significantly sharper for confined water. Moreover, the pressure of the confined liquid remains very close to zero at low temperatures, while the pressure of bulk water is finite and grows continuously even below the boiling point. It should be noted now, that the molecular processes underlying the pressure growth inside CNTs are qualitatively different from the boiling of bulk water, as discussed in detail below. The $T_b$ values listed in Table 1, corresponding to the inflection points in Figure 3, can be called boiling temperatures only nominally. Both the $T_b$ values and the P(T) slopes correlate strongly with the CNT diameter. In particular, $T_b$ is inversely proportional to the diameter.

Further understanding of the effect of the CNT-water interaction on the evaporation of confined water droplets is provided by the data of Figure 4. Here, we use the Lennard-Jones epsilon parameter of the carbon atoms as a tool to control the strength of water attraction to the hydrophobic CNT sidewalls. Focusing on the (31,31) system, we obtained pressure as a function of temperature for a series of diminishing values of epsilon. As should be expected, decreasing the attraction of water to the CNT sidewalls increases pressure. When epsilon approaches zero, the P(T) curve for the confined droplet starts to resembles the bulk data. Nevertheless, even in the absence of the CNT-water attraction, the confined droplet and bulk lines never become identical. The difference in the $T_b$ values is equal to 16 K; $T_b$ = 384 K for confined water in the absence of CNT-water attraction, and $T_b$ = 368 K for bulk TIP4P water.[23] The difference arises solely due to confinement and is smaller for larger tubes. In the limit of small epsilon, the transition from the low pressure region to the region with rapidly rising pressure is smooth, and pressure remains finite even a low temperatures. In contrast for larger epsilon, pressure remains essentially zero at low temperatures, suggesting that water molecules evaporating from the droplet adsorb on the CNT walls.

The fact that water pressure inside CNTs is significantly affected by changes in the van der Waals parameters of CNT sidewalls indicates that pressure can be tuned.[25, 26] Chemical changes made to the outer surface of single-walled CNTs will influence the internal vapor pressure. Relative to the attractive CNT-water interaction, the effect of spatial confinement *per se* on the internal vapor pressure is quite small. Overall, both confinement and the attractive interactions can be used to control quantitatively the evaporation process inside CNTs. Decreasing epsilon by a factor of four is comparable to the change in the CNT diameter from 2.85 to 4.20 nm, compare Figs. 3 and 4. Changing the CNT diameter from 4.20 to 2.85 nm increases $T_b$ by 84 K. Further decrease of the CNT diameter down to 1.49 nm in the (11,11) CNT creates a much more substantial growth in the $T_b$ value, up by 195 K. At the same time, pressure generated in the (11,11) CNT at temperatures above $T_b$ is relatively small, indicating that substantial pressure of water vapor inside CNTs with diameters below 2.5 nm cannot be attained. The large surface-to-volume ratio ensures that adsorption of water on the CNT surface greatly influences thermodynamic properties of liquids confined in small CNTs.

The data presented in Figures 3 and 4 suggest the following scenario for evaporation of water droplets inside CNTs. At temperatures well below $T_b$ for a given tube water molecules with high kinetic energy escape the droplet but are immediately adsorbed on CNT walls. Vapor pressure is not generated. This process can occur at temperatures significantly above the boiling temperature of bulk water. Raising the temperature further increases the kinetic energy of water, which works to overcome the van der Waals attraction releasing water from the CNT walls. As a result, pressure grows rapidly inside CNT.

**The Wetting Transition.** The transport of water molecules from the confined droplet to the CNT walls merits further investigation. Figure 5 shows the water density distribution function calculated along the axial direction of the three CNTs at a series of temperatures. The corresponding bulk data are computed along the direction perpendicular to the water layer. The data for 300K show stable liquid droplets in equilibrium with low density vapor. As the temperature is increased, the droplets disintegrate. The liquid layer representing bulk water disappears around 575 K, corresponding to the critical temperature of the TIP4P water model, Fig. 2. The confined droplets destruct at a significantly lower temperature, $T_c$, around 500 K. Remarkably, unlike most properties of confined media, $T_c$ is independent of the CNT diameter. It is interesting to note that the water droplet confined inside the smallest (11,11) CNT is very mobile. It migrates by about 2 nm to the left within 40 ns of MD simulation. Such behavior can be explained by smoothness of the CNT hydrophobic sidewalls,[27] and a relatively large contact area of the (11,11) CNT and the droplet.

The temperature driven transport of water from the droplet, Fig. 1a, to the CNT walls, Fig. 1c, is an example of a wetting transition: if a fluid does not spread across a surface at a low temperature, then it does so around critical temperature. Our results well agree with the Monte Carlo simulations of water droplets on graphite surface showing the wetting transition at 480 K for a different model of water.[28]

The wetting transition inside CNTs partially filled with water entirely undermines the possibility to use the capillary theory in order to predict the rapid vapor growth of confined water, Figs. 3 and 4. In principle, one can use the Kelvin equation in order to compute the change of the vapor pressure due to the presence of a curved meniscus of water in the CNT capillary. Then one can employ the Clausius-Clapeyron relation in order to estimate the boiling point elevation for the given vapor pressure, relative to the unconfined system. Such calculations provide good agreement with the atomistic simulation for sufficiently large CNTs that are completely filled with water.[23] However, the boiling point elevations on the order of tens of Kelvins predicted by the capillary theory are significantly smaller than the temperatures required for the rapid pressure growth in the current MD calculations, Fig. 3. Such disagreement should be expected,[29] since the meniscus disappears at an elevated temperature, Fig. 1c. The capillary theory neglects the interaction of molecules in the saturated vapor with the walls of the nano-vessel.

Figure 6 depicts the radial density distribution functions calculated perpendicular to the CNT axes for water confined by the three CNTs at 500 K. At this temperature, the droplets have already disintegrated, see Fig. 5. The radial distribution function for the (31,31) CNT shows two distinct density maxima near the opposite sidewalls. The widths of the maxima correspond to about two layers of adsorbed water. The slice across the (31,31) CNT system shown in Fig. 1c gives an atomistic illustration of this situation: water molecules are adsorbed at the CNT walls, and a 1 nm wide cylinder of vacuum is created in the middle of the tube. The CNT-water attraction provides an additional force driving disintegration of the water droplets. It explains why the confined water droplets destruct at a temperature that is lower than the critical temperature of bulk water.

Water molecules inside the smaller (21,21) and (11,11) CNTs are also driven to CNT walls following droplet disintegration at 500 K. However, no vacuum is seen in the middle of the tubes, Fig. 6. Only 7 (in (21,21) CNT) and 3 (in (11,11) CNT) water molecules can be arranged perpendicular to the CNT axis in these tubes. The inner radius of the (21,21) CNT is only 1.3 nm, and dipole-dipole interactions of water molecules adsorbed on the opposite sidewalls are strong. This strong electrostatic interaction stabilizes a continuous distribution of molecules across the smaller tubes. The strong intermolecular interaction of water molecules across the whole width of the two smaller CNTs explains why vapor pressure remains nearly zero to significantly higher temperatures: compare 897 and 702 K for the boiling temperatures in the (11,11) and (21,21) CNTs with 618 K for the (31,31) CNT, Table 1.

**Thermodynamic Considerations.** The information presented above is now sufficient to rationalize the qualitative and quantitative differences in the evaporation of water droplets inside CNTs and in the free state. A free water droplet exists in equilibrium with its vapor pressure. If the volume of the system is fixed as in the current study, vapor pressure grows slowly with increasing temperature below the boiling point and rapidly above the boiling point. Finally, the water droplet disintegrates at the critical temperature. This sequence of events is reversed inside CNTs. The droplet disintegrates while pressure remains nearly zero, and only at a much higher temperature pressure starts growing abruptly and at a rapid pace. These differences arise primarily due to the attractive interaction between the water droplet and CNT walls. The interaction prevents evaporated water molecules from diffusing freely inside the tube. Instead, they adsorb onto CNT sidewalls.

Thermodynamically, one can consider several phases of water in partially filled CNTs, including water droplet (3D liquid), water layer on CNT walls (2D liquid), individual water molecules physisorbed onto CNT walls (2D gas), and confined water vapor (3D gas). The energies and entropies of these phases vary in opposite to each other. In particular, the interaction energy decreases in the above series, since the water-water interaction is stronger than the water-CNT interaction. At the same time, the entropy grows in this series, as water molecules are allowed to explore a larger fraction of space inside CNTs. Raising temperature increases the importance of the entropy term in the thermodynamic free energy of the system. As a result, the system undergoes a sequence of transitions from 3D liquid to 2D liquid (wetting transition), to 2D gas (not seen in our calculations due to high concentration of water molecules), and then to 3D gas (evaporation).

The CNT-water interaction is responsible for lowering the droplet disintegration temperature from the critical temperature of 575 K for bulk TIP4P water down to 500 K. The value of 500 K is determined by the strength of the interaction and is largely independent of the CNT diameter. Once water transforms from a droplet into a film adsorbed on CNT walls, vapor pressure can be increased only when temperature becomes sufficiently high to overcome both the adsorption energy and the intermolecular interactions of water molecules within this layer. The desorption temperature increases in small CNTs, because additional energy is required to overcome the electrostatic interaction between water layers adsorbed on the opposite walls.

**Experimental Implications.** The temperature dependence of the water droplet behavior predicted by our simulations can be detected experimentally, for instance, using transmission electron microscopy (TEM).[29] Our results indicate that it is particularly important in such experimental studies to achieve good control over CNT quality and functionalization, since the presence of defects and different functional groups on CNT walls can alter the strength of the CNT-water interaction and change the quantitative details of the considered phenomena. At the same time, the qualitative sequence of events predicted in our simulation should remain robust to variations in CNT samples.

The sharp growth of vapor pressure inside CNTs above a certain temperature, whose value can be controlled by CNT diameter and functionalization, can prove advantageous in CNT applications for drug delivery. The threshold temperature values may turn out too high for this purpose. In such case, one can use other solvents with lower boiling temperatures and weaker intermolecular interactions, such as carbon dioxide or a carbon dioxide/water mixture, so-called carbonated water. The desired increase in the vapor pressure can be achieved at a significantly lower temperature for CNTs filled with water completely[23] compared to the CNT/droplet systems considered in the current work.

**CONCLUSIONS**

To recapitulate, MD simulations of water droplets inside the (11,11), (21,21) and (31,31) CNTs have been carried out over a wide range of temperatures. Under confinement, water droplets disintegrate at a lower temperature than in the bulk phase; however, little vapor pressure is generated. Sharp growth of vapor pressure inside CNTs is achieved by heating the system substantially above the droplet disintegration temperature. This situation is in contrast to the behavior of free water droplets, which produce high vapor pressure at the boiling temperature that is much smaller than the critical temperature, $T_c$, required for droplet disintegration. The differences in the evaporation mechanisms of the free and confined water droplets are rationalized by the CNT-water interaction. First, the droplet transforms into a film of water molecules adsorbed at CNT walls. The temperature of this process is independent of the CNT diameter, which is quite unusual for properties of confined liquids. The droplet disintegration temperature is determined by the strength of the CNT-water interaction. Second, at a substantially higher temperature water abruptly generates strong vapor pressure. This temperature does depend on the CNT diameter for small CNTs, since it is determined not only by the CNT-water interaction, but also by the interaction between water layers adsorbed on the opposite sides of the CNTs. The sharp increase of pressure inside CNTs upon heating, and the control over the temperature of this transition provided by the CNT diameter and sidewall functionalization, can be utilized in therapeutic applications of CNTs.

**METHODS**

The simulations were performed with systems consisting of a water droplet encapsulated within a multi-walled CNT, Figure 1a. Three co-axial CNTs were used to prevent interaction of inner and outer water molecules. Guided by the 0.34 nm experimental interlayer distance, the multi-walled CNTs were selected such that an (n,n) inner-most CNT defined in Table 1 was encapsulated by the (n+5, n+5) and (n+10, n+10) outer CNTs. A water layer in equilibrium with its vapor pressure in a series of rectangular boxes was used to represent the properties of unconfined water, system 1 in Table 1.

The simulated CNTs were treated as flexible non-polarizable objects. The potentials for bonds, angles, dihedrals and pairwise 1-4 carbon-carbon Lennard-Jones interactions within the CNTs were represented by the AMBER force field (FF).[30] Water molecules were represented by the TIP4P four-site potential.[31] Among the incredibly diverse set of water FFs, TIP4P is most successful in reproducing the liquid/vapor phase transition temperature, Figure 2. The electrostatic interaction energy for pairs of solvent molecules was calculated within the 1.3 nm cut-off radius using the reaction-field-zero technique.[32] The Lennard-Jones (12,6) interactions were computed using the shifted force method,[32] with the switch region from 1.1 to 1.2 nm. The Lorenz-Berthelot combination rules were used to derive the cross-term Lennard-Jones parameters. The selection of the interaction potentials for the CNT-water systems followed the well-tested practice.[33]

The trajectories were propagated using GROMACS.[32] The leap-frog integration algorithm with 0.002 ps time-step was applied. The simulations started with empty, multi-walled CNTs surrounded with liquid water. A 5,000 ps MD simulation in the NPT ensemble allowed water to enter CNTs and achieve relaxed configurations.[34] The thermostat of Bussi et al.,[35] often referred to as "V-rescale", with the response time of 0.5 ps was used to maintain temperature at 300 K. The Parrinello-Rahman barostat[36] with the time constant of 4.0 ps was used to maintain pressure at 1 bar. Following initial equilibration, all water molecules outside CNTs were deleted. Water droplets were created by removing 3.5 nm of confined water from each end of the CNTs, Figure 1a.

Simulated annealing was carried out by gradually heating the systems from 300 to 1000 K with at the rate of 1.0 degree per 200 ps. To ensure that the simulated water structure had sufficient time to adapt to the constantly increasing temperature, NVT runs were performed at several temperatures selected arbitrarily between 300 and 1000 K. The vapor pressure did not change during these test runs.

The pressure in the confined systems was defined as $\mathbf{P} = 2 \cdot (\mathbf{E}_{kin} - \mathbf{G})/V$, where V is system volume, and $\mathbf{P}$, $\mathbf{E}_{kin}$ and $\mathbf{G}$ are pressure, kinetic energy, and virial tensors, respectively. The virial tensor is given by $\mathbf{G} = -\frac{1}{2} \sum_{i<j} \mathbf{r}_{ij} \otimes \mathbf{F}_{ij}$. The summations in the above equations involved pairs of atoms belonging only to water molecules, excluding CNT atoms. For isotropic systems, pressure is derived as $P = trace(\mathbf{P})/3$. For the confined droplets, pressure was defined by its component along the CNT axis.

**Acknowledgment**

The authors are grateful to H. Jaeger and Yu. Pereverzev for comments on the manuscript. The research was supported by NSF grant CHE-1050405.

**Table 1**. Simulated systems. Each system consists of a water droplet inside a CNT, see Fig. 1a. The CNTs are 10 nm long. System 1 represents bulk TIP4P water and includes a water layer in a rectangular box, as detailed in the text. $T_b$ and $T_c$ correspond to the steep rise of the vapor pressure and droplet disintegration. For bulk water, these are the boiling and critical temperatures, Fig. 2. The order of the temperatures is reversed for confined droplets, for reasons described in the text.

| System | CNT | $d_{CNT}$, nm | $N_{waters}$ | $T_b$, K | $T_c$, K |
|--------|---------|------|------|-----|-----|
| 1      | —       | —    | 1728 | 368 | 575 |
| 2      | (11,11) | 1.49 | 89   | 897 | 490 |
| 3      | (21,21) | 2.85 | 462  | 702 | 500 |
| 4      | (31,31) | 4.20 | 1112 | 618 | 490 |

**FIGURE CAPTIONS**

**Figure 1**. (Color online) (31,31) CNT containing TIP4P water. **a)** At a low temperature, T=300 K, water exists as a droplet. **b)** At a high temperature, T=650 K, evaporated water is distributed uniformly throughout the tube. **c)** At an intermediate temperature, T=500 K, water adsorbs onto CNT sidewalls.

**Figure 2**. (Color online) Pressure *vs.* temperature for bulk water obtained for system 1, Table 1, placed in boxes with the same cross-section and different lengths: 10 nm (red solid), 20 nm (green dashed) and 30 nm (blue dash-dotted). Each system contains 1728 water molecules. The arrows show the boiling and critical temperatures, 368 K and 575 K, respectively. Above its critical point, water cannot be condensed at any temperature; therefore the three lines corresponding to different system volumes diverge.

**Figure 3**. Pressure *vs.* temperature for bulk and confined water droplets. Pressure rises rapidly above a certain threshold temperature, nominally boiling temperature, which is different for each system depending on the strength of the confinement effect.

**Figure 4**. Pressure *vs.* temperature in the (31,31) CNT system with different degrees of attraction of water molecules to CNT carbon atoms. The attraction strength was varied by changing the epsilon value in the Lennard-Jones equation. As epsilon decreases, the data approach those for bulk water, Fig. 3.

**Figure 5**. (Color online) Water density distribution for free liquid film and inside the (31,31), (21,21) and (11,11) CNTs computed along the tube axis at several temperatures. In bulk, the density becomes uniform above the critical temperature of 575K. Under confinement, the droplet disintegrates at a lower temperature of about 500K, independent of the CNT diameter.

**Figure 6**. Radial density distributions of water confined inside the (31,31), (21,21) and (11,11) CNTs at 500 K. Water molecules tend to adsorb onto CNT walls, as illustrated in Fig. 1c.

# REFERENCES

1. Chen, X.; Cao, G. X.; Han, A. J.; Punyamurtula, V. K.; Liu, L.; Culligan, P. J.; Kim, T.; Qiao, Y., Nanoscale Fluid Transport: Size and Rate Effects. *Nano Lett.* **2008**, *8*(9), 2988-2992.
2. Das, A.; Jayanthi, S.; Deepak, H.; Ramanathan, K. V.; Kumar, A.; Dasgupta, C.; Sood, A. K., Single-File Diffusion of Confined Water inside SWNTs: An NMR Study. *ACS Nano* **2010**, *4*(3), 1687-1695.
3. Han, S. H.; Choi, M. Y.; Kumar, P.; Stanley, H. E., Phase Transitions in Confined Water Nanofilms. *Nature Physics* **2010**, *6*(9), 685-689.
4. Koga, K.; Gao, G. T.; Tanaka, H.; Zeng, X. C., Formation of Ordered Ice Nanotubes inside Carbon Nanotubes. *Nature* **2001**, *412*(6849), 802-805.
5. Chaban, V. V.; Savchenko, T. I.; Kovalenko, S. M.; Prezhdo, O. V., Heat-Driven Release of a Drug Molecule from Carbon Nanotubes: A Molecular Dynamics Study. *J. Phys. Chem. B* **2010**, *114*(42), 13481-13486.
6. Waghe, A.; Rasaiah, J. C.; Hummer, G., Filling and Emptying Kinetics of Carbon Nanotubes in Water. *J. Chem. Phys.* **2002**, *117*(23), 10789-10795.
7. Rasaiah, J. C.; Garde, S.; Hummer, G., Water in Nonpolar Confinement: From Nanotubes to Proteins and Beyond. *Annu. Rev. Phys. Chem.* **2008**, *59*, 713-740.
8. Hummer, G., Water, Proton, and Ion Transport: From Nanotubes to Proteins. *Mol. Phys.* **2007**, *105*(2-3), 201-207.
9. Mashl, R. J.; Joseph, S.; Aluru, N. R.; Jakobsson, E., Anomalously Immobilized Water: A New Water Phase Induced by Confinement in Nanotubes. *Nano Lett.* **2003**, *3*(5), 589-592.
10. Kolesnikov, A. I.; Zanotti, J. M.; Loong, C. K.; Thiyagarajan, P.; Moravsky, A. P.; Loutfy, R. O.; Burnham, C. J., Anomalously Soft Dynamics of Water in a Nanotube: A Revelation of Nanoscale Confinement. *Phys. Rev. Lett.* **2004**, *93*(3), 035503.
11. Kalugin, O. N.; Chaban, V. V.; Loskutov, V. V.; Prezhdo, O. V., Uniform Diffusion of Acetonitrile inside Carbon Nanotubes Favors Supercapacitor Performance. *Nano Lett.* **2008**, *8*(8), 2126-2130.
12. Giovambattista, N.; Rossky, P. J.; Debenedetti, P. G., Phase Transitions Induced by Nanoconfinement in Liquid Water. *Phys. Rev. Lett.* **2009**, *102*(5), 050603
13. Jiang, J. W.; Sandler, S. I.; Smit, B., Capillary Phase Transitions of N-Alkanes in a Carbon Nanotube. *Nano Lett.* **2004**, *4*(2), 241-244.
14. Kostarelos, K.; Bianco, A.; Prato, M., Promises, Facts and Challenges for Carbon Nanotubes in Imaging and Therapeutics. *Nature Nanotech.* **2009**, *4*(10), 627-633.
15. Kim, C.; Cho, E. C.; Chen, J. Y.; Song, K. H.; Au, L.; Favazza, C.; Zhang, Q. A.; Cobley, C. M.; Gao, F.; Xia, Y. N*., et al.*, In Vivo Molecular Photoacoustic Tomography of Melanomas Targeted by Bioconjugated Gold Nanocages. *ACS Nano* **2010**, *4*(8), 4559-4564.
16. Zheng, G. F.; Gao, X. P. A.; Lieber, C. M., Frequency Domain Detection of Biomolecules Using Silicon Nanowire Biosensors. *Nano Lett.* **2010**, *10*(8), 3179-3183.
17. Hong, S. Y.; Tobias, G.; Al-Jamal, K. T.; Ballesteros, B.; Ali-Boucetta, H.; Lozano-Perez, S.; Nellist, P. D.; Sim, R. B.; Finucane, C.; Mather, S. J*., et al.*, Filled and Glycosylated Carbon Nanotubes for in Vivo Radioemitter Localization and Imaging. *Nature Mater.* **2010**, *9*(6), 485-490.
18. Wang, M. Y.; Yu, S. N.; Wang, C. A.; Kong, J. L., Tracking the Endocytic Pathway of Recombinant Protein Toxin Delivered by Multiwalled Carbon Nanotubes. *ACS Nano* **2010**, *4*(11), 6483-6490.
19. Zavaleta, C.; de la Zerda, A.; Liu, Z.; Keren, S.; Cheng, Z.; Schipper, M.; Chen, X.; Dai, H.; Gambhir, S. S., Noninvasive Raman Spectroscopy in Living Mice for Evaluation of Tumor Targeting with Carbon Nanotubes. *Nano Lett.* **2008**, *8*(9), 2800-2805.
20. Gannon, C. J.; Cherukuri, P.; Yakobson, B. I.; Cognet, L.; Kanzius, J. S.; Kittrell, C.; Weisman, R. B.; Pasquali, M.; Schmidt, H. K.; Smalley, R. E*., et al.*, Carbon Nanotube-Enhanced Thermal Destruction of Cancer Cells in a Noninvasive Radiofrequency Field. *Cancer* **2007**, *110*(12), 2654-2665.

21. Luer, L.; Lanzani, G.; Crochet, J.; Hertel, T.; Holt, J.; Vardeny, Z. V., Ultrafast Dynamics in Metallic and Semiconducting Carbon Nanotubes. *Phys. Rev. B* **2009**, *80*(20), 205411.
22. Nelson, T. R.; Chaban, V. V.; Kalugin, O. N.; Prezhdo, O. V., Vibrational Energy Transfer between Carbon Nanotubes and Liquid Water: A Molecular Dynamics Study. *J. Phys. Chem. B* **2010**, *114*(13), 4609-4614.
23. Chaban, V. V.; Prezhdo, O. V., Water Boiling inside Carbon Nanotubes: Toward Efficient Drug Release. *ACS Nano* **2011**, *5*(7), 5647-5655.
24. Vega, C.; Abascal, J. L. F.; Nezbeda, I., Vapor-Liquid Equilibria from the Triple Point up to the Critical Point for the New Generation of TIP4P-like Models: TIP4P/EW, TIP4P/2005, and TIP4P/Ice. *J. Chem. Phys.* **2006**, *125*(3), 034503.
25. Ueta, A.; Tanimura, Y.; Prezhdo, O. V., Distinct Infrared Spectral Signatures of the 1,2-and 1,4-Fluorinated Single-Walled Carbon Nanotubes: A Molecular Dynamics Study. *J. Phys. Chem. Lett.* **2010**, *1*(9), 1307-1311.
26. Yarotski, D. A.; Kilina, S. V.; Talin, A. A.; Tretiak, S.; Prezhdo, O. V.; Balatsky, A. V.; Taylor, A. J., Scanning Tunneling Microscopy of DNA-Wrapped Carbon Nanotubes. *Nano Lett.* **2009**, *9*(1), 12-17.
27. Hummer, G.; Rasaiah, J. C.; Noworyta, J. P., Water Conduction through the Hydrophobic Channel of a Carbon Nanotube. *Nature* **2001**, *414*(6860), 188-190.
28. Zhao, X. C., Wetting Transition of Water on Graphite: Monte Carlo Simulations. *Phys. Rev. B* **2007**, *76*(4), 041402.
29. Mattia, D.; Gogotsi, Y., Review: Static and Dynamic Behavior of Liquids inside Carbon Nanotubes. *Microfluid Nanofluid* **2008**, *5*(3), 289-305.
30. Yang, L. J.; Luo, R., Amber United-Atom Force Field. *Biophys. J.* **2005**, *88*(1), 512a-512a.
31. Ferrario, M.; Tani, A., A Molecular-Dynamics Study of the Tip4p Model of Water. *Chem. Phys. Lett.* **1985**, *121*(3), 182-186.
32. Hess, B.; Kutzner, C.; van der Spoel, D.; Lindahl, E., Gromacs 4: Algorithms for Highly Efficient, Load-Balanced, and Scalable Molecular Simulation. *J. Chem. Theory Comput.* **2008**, *4*(3), 435-447.
33. Alexiadis, A.; Kassinos, S., Molecular Simulation of Water in Carbon Nanotubes. *Chem. Rev.* **2008**, *108*(12), 5014-5034.
34. Chaban, V., Should Carbon Nanotubes Be Degasified before Filling? *Chem. Phys. Lett.* **2010**, *500*(1-3), 35-40.
35. Bussi, G.; Donadio, D.; Parrinello, M., Canonical Sampling through Velocity Rescaling. *J. Chem. Phys.* **2007**, *126*(1), 014101.
36. Parrinello, M.; Rahman, A., Polymorphic Transitions in Single-Crystals - a New Molecular-Dynamics Method. *J. Appl. Phys.* **1981**, *52*(12), 7182-7190.

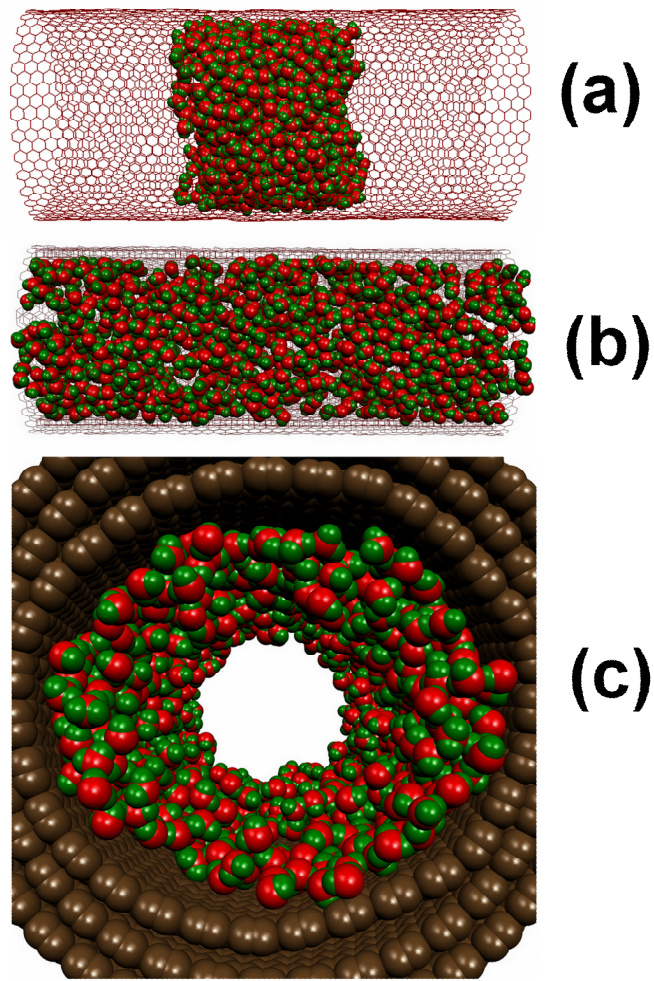

FIGURE 2

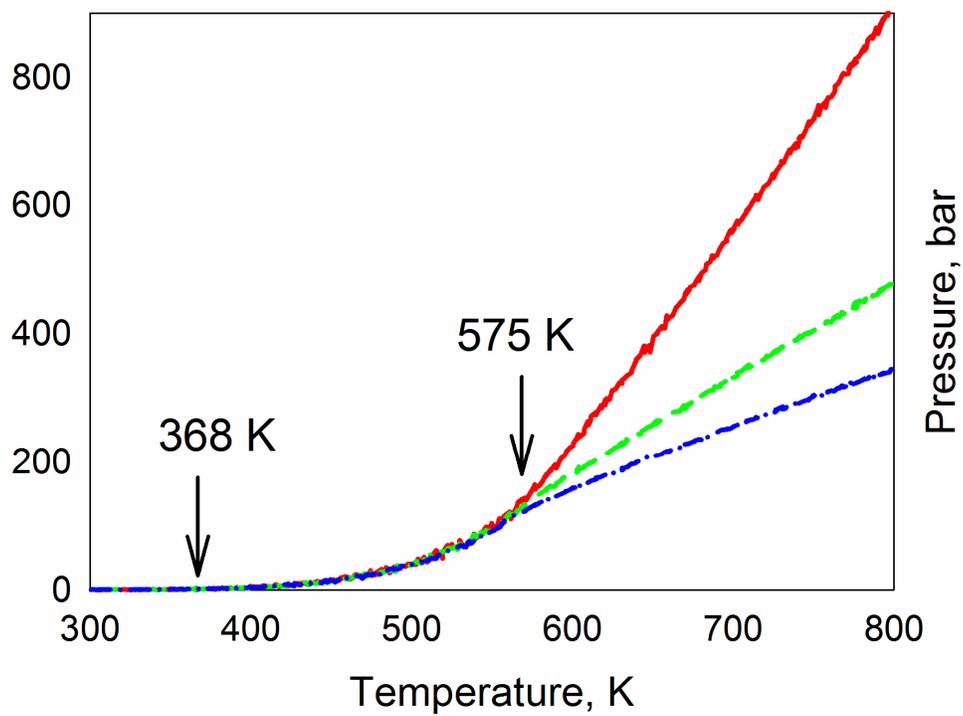

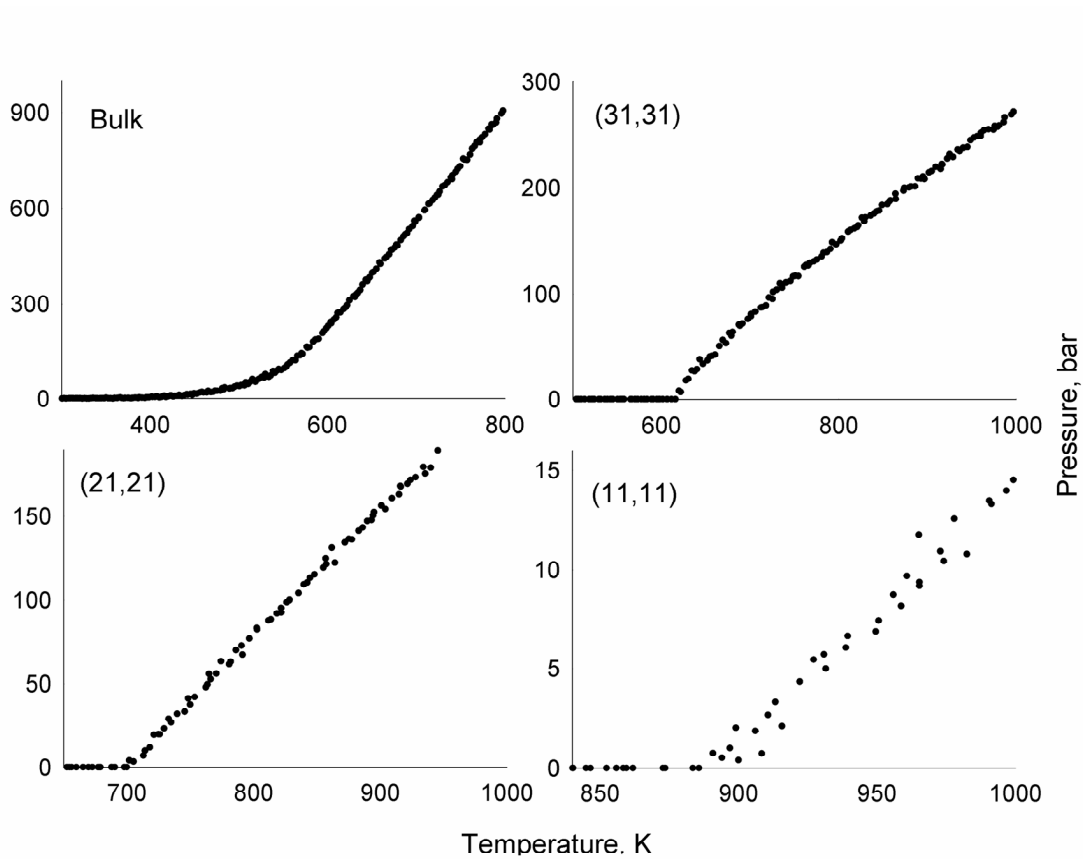

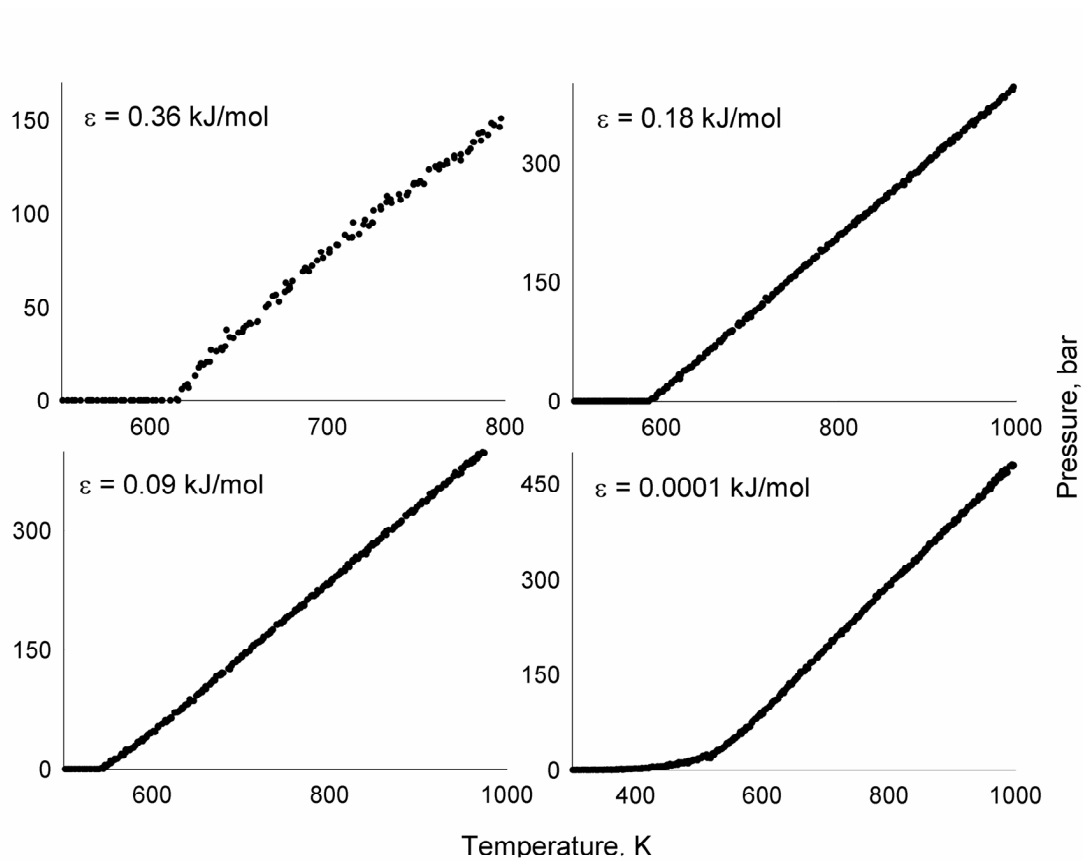

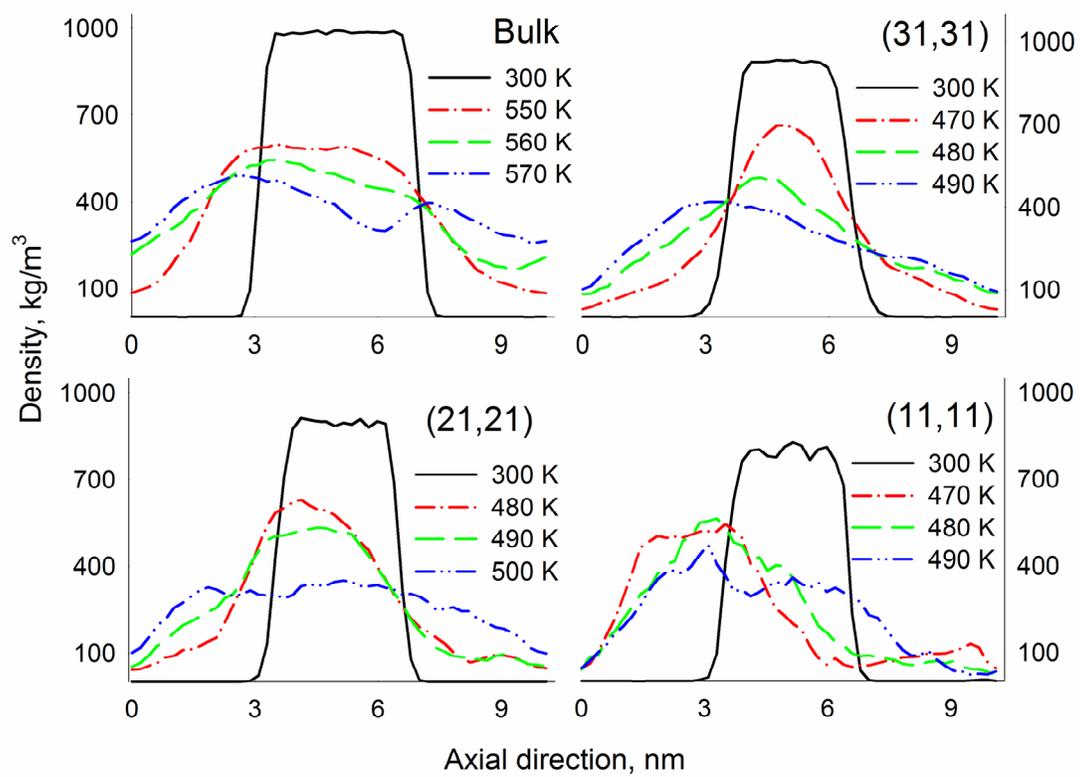

FIGURE 6FIGURE 6

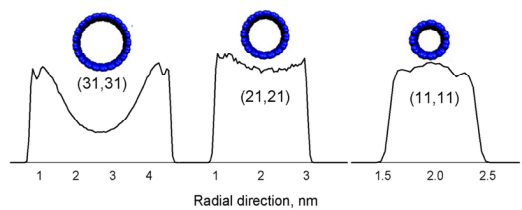

FIGURE 6

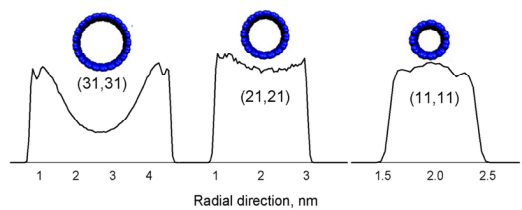